\documentclass{aadebug}

\corr{0309027}{103}

\usepackage{graphicx} 
\usepackage{epsfig}
\usepackage{url}

\makeatletter
\def\ifundefined{\@ifundefined}
\makeatother

\newcommand\diota{{\em DIOTA}}

\begin{document}

\runningheads{Jonas Maebe, Koen De Bosschere}{Instrumenting self-modifying code}

\title{Instrumenting self-modifying code}

\author{
Jonas~Maebe\addressnum{1},
Koen~De~Bosschere\addressnum{1}\comma\extranum{1}
}

\address{1}{
ELIS,
Ghent University,
Sint-Pietersnieuwstraat 41,
9000 Gent,
Belgium
}

\extra{1}{E-mail: \{jmaebe,kdb\}@elis.UGent.be}

\pdfinfo{
/Title (Instrumenting self-modifying code)
/Author (Jonas Maebe, Koen De Bosschere)
}

\csname @ifundefined\endcsname{pdfoutput}{
\DeclareGraphicsExtensions{.eps}
}{
\DeclareGraphicsExtensions{.pdf}
}

\begin{abstract} Adding small code snippets at key points to existing code fragments is called instrumentation. It is an established technique to debug certain otherwise hard to solve faults, such as memory management issues and data races. Dynamic instrumentation can already be used to analyse code which is loaded or even generated at run time. With the advent of environments such as the Java Virtual Machine with optimizing Just-In-Time compilers, a new obstacle arises: self-modifying code. In order to instrument this kind of code correctly, one must be able to detect modifications and adapt the instrumentation code accordingly, preferably without incurring a high penalty speedwise. In this paper we propose an innovative technique that uses the hardware page protection mechanism of modern processors to detect such modifications. We also show how an instrumentor can adapt the instrumented version depending on the kind of modificiations as well as an experimental evaluation of said techniques. \end{abstract}


\section{Introduction}

Instrumentation is a technique whereby existing code is modified in order to observe or modify its behaviour. It has a lot of different applications, such as profiling, coverage analysis and cache simulations. One of its most interesting features is however the ability to perform automatic debugging, or at least assist in debugging complex programs. After all, instrumentation code can intervene in the execution at any point and examine the current state, record it, compare it to previously recorded information and even modify it.

Debugging challenges that are extremely suitable for analysis through instrumentation include data race detection~\cite{mr-tocs,wbt-acm} and memory management checking~\cite{valgrind}. These are typically problems that are very hard to solve manually. However, since they can be described perfectly using a set of rules (e.g.\ the memory must be allocated before it is accessed, or no two threads must write to the same memory location without synchronising), they are perfect candidates for automatic verification. Instrumentation provides the necessary means to insert this verification code with little effort on the side of the developer.

The instrumentation can occcur at different stages of the compilation or execution process. When performed prior to the execution, the instrumentation results in changes in the object code on disk, which makes them a property of a program or library. This is called static instrumentation.  If the addition of instrumentation code is postponed until the program is loaded in memory, it becomes a property of an execution. In this case, we call it dynamic instrumentation.

Examples of stages where static instrumentation can be performed are directly in the source code~\cite{insight}, in the assembler output of the compiler~\cite{MPTRACE}, in the compiled objects or programs (e.g.\ EEL~\cite{EEL}, ATOM~\cite{ATOM}, alto~\cite{alto}). The big advantage of static instrumentation is that it must be done only once, after which one can perform several executions without having to reinstrument the code every time. This means that the cost of instrumenting the code can be relatively high without making such a tool practically unusable.

The larges disadvantage of static instrumentation is that it requires a complex analysis of the target application to detect all possible execution paths, which is not always possible. Additionally, the user of a static instrumentation tool must know which libraries are loaded at run time by programs he wants to observe, so that he can provide instrumented versions of those. Finally, every time a new type of instrumentation is desired, the application and its libraries must be reinstrumented.

Most of the negative points of static instrumentation are solved in its dynamic counterpart. In this case, the instrumentation is not performed in advance, but gradually at run time as more code is executed. Since the instrumentation can continue while the program is running, no prior analysis of all possible execution paths is required. It obviously does mean that the instrumentation must be redone every time the program is executed. This is somewhat offset by having to instrument only the part of the application and its libraries that is covered by a particular execution though. One can even apply dynamic optimization techniques~\cite{jm024} to further reduce this overhead.

When using dynamic instrumentation, the code on disk is never modified. This means that a single uninstrumented copy of an application and its libraries suffices when using this technique, no matter how many different types of instrumentation one wants to perform. Another consequence is that the code even does not have to exist on disk. Indeed, since the original code is read from memory and can be instrumented just before it is executed, even dynamically loaded and generated code pose no problems. However, when the program starts modifying this code, the detection and handling of these modifications is not possible using current instrumentation techniques.

Yet, being able to instrument self-modifying code becomes increasingly interesting as run time systems that exhibit such behaviour gain more and more popularity. Examples include Java Virtual Machines, the .NET environment and emulators with embedded Just-in-Time compilers in general. These environments often employ dynamic optimizing compilers which continuously change the code in memory, mainly for performance reasons.

Instrumenting the programs running in such an environment is often very easy. After all, the dynamic compiler or interpreter that processes said programs can do the necessary instrumentation most of the time. On the other hand, observing the interaction of the environments themselves with the applications on top and with the underlying operating system is much more difficult.  Nevertheless, this ability is of paramount importance when analysing the total workload of a system and debugging and enhancing these virtual machines.

Even when starting from a system that can already instrument code on the fly, supporting self-modifying code is a quite complex undertaking. First of all, the original program code must not be changed by the instrumentor, since otherwise the program's own modifications may conflict with these changes later on. Secondly, the instrumentor must be able to detect changes performed by the program before the modified code is executed, so that it can reinstrument this code in a timely manner. Finally, the reinstrumentation itself must take into account that an instruction may be changed using multiple write operations, so it could be invalid at certain points in time.

In this paper we propose a novel technique that can be used to dynamically instrument self-modifying code with an acceptable overhead. We do this by using the hardware page protection facilities of the processor to mark pages that contain code which has been instrumented as read-only. When the program later on attempts to modify instrumented code, we catch the resulting protection faults which enables us to detect those changes and act accordingly. The described method has been experimentally evaluated using the \diota{} ({D}ynamic {I}nstrumentation, {O}ptimization and {T}ransformation of {A}pplications~\cite{wbt-diota}) framework on the Linux/x86 platform by instrumenting a number of JavaGrande~\cite{javagrande} benchmarks running in the Sun 1.4.0 Java Virtual Machine.

The paper now proceeds with an overview of dynamic instrumentation in general and \diota{} in particular. Next, we show how the detection of modified code is performed and how to reinstrument this code. We then present some experimental results of our implementation of the described techniques and wrap up with the conclusions and our future plans.


\section{Dynamic instrumentation}


\subsection{Overview}

Dynamic instrumentation can be be done in two ways. One way is modifying the existing code, e.g.\ by replacing instructions with jumps to routines which contain both instrumentation code and the replaced instruction~\cite{paradyn}. This technique is not very usable on systems with variable-length instructions however, as the jump may require more space than the single instruction one wants to replace. If the program later on transfers control to the second instruction that has been  replaced, it will end up in the middle of this jump instruction. The technique also wreaks havoc in cases of data-in-code or code-in-data, as modifying the code will cause modifications to the data as well.

The other approach is copying the original code into a separate memory block (this is often called {\em cloning}) and adding instrumentation code to this copy~\cite{jm024,jm025,wbt-diota}. This requires special handling of control-flow instructions with absolute target addresses, since these addresses must be relocated to the instrumented version of the code. On the positive side, 
data accesses still occur correctly without any special handling, even in data-in-code situations.

The reason is that when the code is executed in the clone, only the program counter (PC) has a different value in an instrumented execution compared to a normal one. This means that when a program uses non-PC-relative addressing modes for data access, these addresses still refer to the original, unmodified copy of the program or data. PC-relative data accesses can be handled at instrumentation time, as the instrumentor always knows the address of the instruction it is currently instrumenting. This way, it can replace PC-relative memory accesses with absolute memory accesses based on the value the PC would have at that time in a uninstrumented execution.

\subsection{\diota{}}

\begin{figure}
\center{
 \includegraphics[width=8.5 cm]{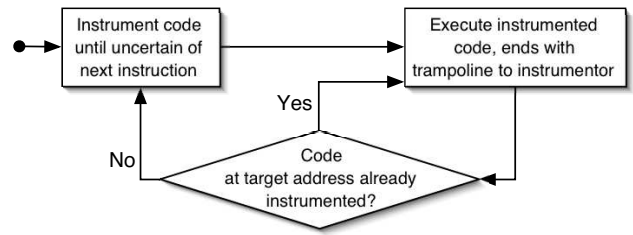}
 \caption{Dynamic instrumentation the \diota{} way}
 \label{fig:diota_operation}
}
\end{figure}

\diota{} uses the cloning technique together with a cache that keeps track of already translated instruction blocks. It is implemented as a shared library and thus resides in the same address space as the program it instruments. By making use of the {\tt LD\_PRELOAD} environment variable under Linux, the dynamic linker ({\tt ld.so}) can be forced to load this library, even though an application is not explicitly linked to it. The {\tt init} routines of all shared libraries are executed before the program itself is started, providing \diota{} an opportunity to get in control.

As shown in Figure \ref{fig:diota_operation}, the instrumentation of a program is performed gradually. First, the instructions at the start of the program are analysed and then copied, along with the desired instrumentation code, to the {\em clone} (a block of memory reserved at startup time, also residing in the program's address space). During this process, direct jumps and calls are followed to their destination. The instrumentation stops when an instruction is encountered of which the destination address cannot be determined unequivocally, such as an indirect jump.

At this point, a {\em trampoline} is inserted in the clone. This is a small piece of code which will pass the actual target address to \diota{} every time the corresponding original instruction would be executed. For example, in case of a jump with the target address stored in a register, the trampoline will pass the value of that specific register to \diota{} every time it is executed. When \diota{} is entered via such a trampoline, it will check whether the code at the passed address has already been instrumented. If that is not the case, it is instrumented at that point. Next, the instrumented version is executed.

\begin{figure} \center{ \includegraphics[width=8.5 cm]{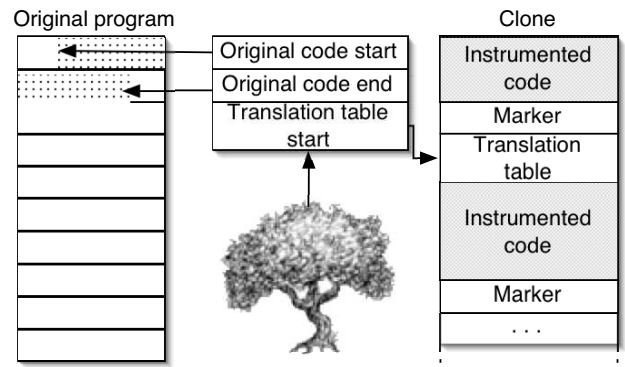} \caption{Data structures used by \diota{}} \label{fig:translationtable} } \end{figure}

Figure \ref{fig:translationtable} shows how \diota{} keeps track of which instructions it has already instrumented and where the instrumented version can be found. A marker consisting of illegal opcodes is placed after every block of instrumented code (aligned to a 4-byte boundary), followed by the translation table. Such a translation table starts with two 32 bit addresses: the start of the block in the original code and its counterpart in the clone. Next, pairs of 8 bit offsets between two successive instructions in the respective blocks are stored, with an escape code to handle cases where the offset is larger than 255 bytes (this can occur because we follow direct calls and jumps to their destination).

In addition to those tables, an AVL tree is constructed. The keys of its elements are the start and stop addresses of the blocks of original code that have been instrumented. The values are the start addresses of the translation tables of the corresponding instrumented versions. Every instruction is instrumented at most once, so the keys never overlap. This means that finding the instrumented version of an instruction boils down to first searching for its address in the AVL tree and if found, walking the appropriate translation table. To speed up this process, a small hash table is used which keeps the results of the latest queries.

A very useful property of this system is that it also works in reverse: given the address of an instrumented instruction, it is trivial to find the address of corresponding original instruction. First, the illegal opcodes marker is sought starting from the queried address and next the table is walked just like before until the appropriate pair is found. This ability of doing two-way translations is indispensable for the self-modifying code support and proper exception handling.

Since the execution is followed as it progresses, code-in-data and code loaded or generated at run time can be handled without any problems. When a trampoline passes an address to \diota{} of code it has not yet instrumented, it will simply instrument it at that time. It is irrelevant where this code is located, when it appeared in memory and whether or not it doubles as data

\diota{} has several modes of operation, each of which can be used separately, but most can be combined as well.
Through the use of so-called backends, the different instrumentation modes can be activated and the instrumentation parameters can be modified. These backends are shared libraries that link against \diota{} and which can ask to intercept calls to arbitrary dynamically linked routines based on name or address, to have a handler called whenever a memory access occurs, when a basic block completes or when a system call is performed (both before and after the system call, with the ability to modify its parameters or return value). Several backends can be used at the same time.

Other features of the \diota{} framework include the ability to handle most extensions to the 80x86 ISA (such as MMX, 3DNow! and SSE) and an extensible and modular design that allows easy implementation of additional backends and support for newly introduced instructions. This paper describes the support for instrumenting self-modifying code in \diota{}. For other technical details about \diota{} we refer to \cite{wbt-diota}.

\subsection{Exception handling}

An aspect that is of paramount importance to the way we handle self-modifying code, is the handling of exceptions (also called signals under Linux). The next section will describe in more detail how we handle the self-modifying code, but since it is based on marking the pages containing code that has been instrumented as read-only, it is clear that every attempt to modify such code will cause a protection fault (or {\em segmentation fault}) exception.

These exceptions and those caused by other operations must be properly distinguished, in order to make sure that the program still receives signals which are part of the normal program execution while not noticing the other ones. This is especially important since the Java Virtual Machine that we used to evaluate our implementation uses signals for inter-thread communication.

When a program starts up, each signal gets a default handler from the operating system. If a program wants to do something different when it receives a certain signal, it can install a signal handler by performing a system call. This system call gets the signal number and the address of the new handler as arguments. Since we want to instrument these user-installed handlers, we have to intercept these system calls.

This can be achieved by registering a system call analyser routine with \diota{}. This instructs \diota{} to insert a call to this routine after every system call in the instrumented version of the program. If such a system call successfully installed a new signal handler, the analyser records this handler and then installs a \diota{} handler instead.

Next, when a signal is raised, \diota{}'s handler is activated. One of the arguments passed to a signal handler contains the contents of all processor registers at the time the signal occurred, including those of the instruction pointer register. Since the program must not be able to notice it is being instrumented by looking at at that value, it is translated from a clone address to an original program address using the translation tables described previously. Finally, the handler is executed under control of \diota{} like any other code fragment.

Once the execution arrives at the {\tt sig\_return} or {\tt sig\_rt\_return} system call that ends this signal's execution, \diota{} replaces the instruction pointer in the signal context again. If the code at that address is not yet instrumented, the instruction pointer value in the context is replaced with the address of a trampoline which will transfer control back to \diota{} when returning from the signal's execution. Otherwise, the clone address corresponding to the already instrumented version is used.

\section{Detecting modifications}

Dynamically generated and loaded code can already be handled by a number of existing instrumentors~\cite{jm024,wbt-diota}. The extra difficulty of handling self-modifying code is that the instrumentation engine must be able to detect modifications to the code, so that it can reinstrument the new code. Even the reinstrumenting itself is not trivial, since a program may modify an instruction by performing two write operations, which means the intermediate result could be invalid.

There are two possible approaches for dealing with code changes. One is to detect the changes as they are made, the other is to check whether code has been modified every time it is executed. Given the fact that in general code is modified far less than it is executed, the first approach was chosen. The hardware page protection facilities of the processor are used to detect the changes made page. Once a page contains code that has been instrumented, it will be  write-protected. The consequence is that any attempt to modify such code will result in a segmentation fault. An exception handler installed by \diota{} will intercept these signals and take the appropriate action.

\begin{figure}
\center{
  \includegraphics[width=8.5 cm]{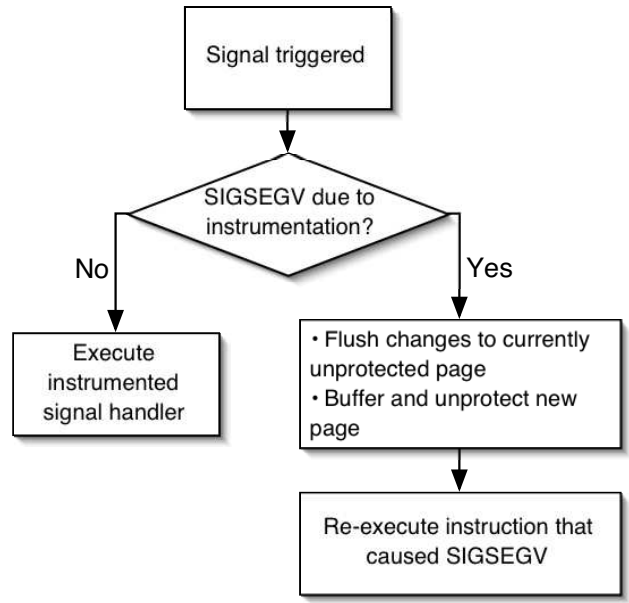}
  \caption{Exception handling in the context of self-modifying code support}
  \label{fig:exceptions}
}
\end{figure}

Since segmentation faults must always be caught when using our technique to support self-modifying code, \diota{} installs a dummy handler at startup time and whenever a program installs the default system handler for this signal (which simply terminates the process if such a signal is raised), or when it tries to ignore it. Apart from that, no changes to the exception handling support of \diota{} have been made, as shown in Figure \ref{fig:exceptions}.

Whenever a protection fault occurs due to the program trying to modify some previously instrumented code, a naive implementation could unprotect the relevant page, perform the required changes to the instrumented code inside the signal handler, reprotect the page and continue the program at the next instruction. There are several problems with this approach however:

\begin{itemize}
\item On a CISC architecture, most instructions can access memory, so decoding the instruction that caused the protection fault (to perform the change that caused the segmentation fault in the handler) can be quite complex.
\item It is possible that an instruction is modified by means of more than one memory write operation. Trying to reinstrument after the first write operation may result in encountering an invalid instruction.
\item In the context of a {J}{i}{T}-compiler, generally more than one write operation occurs to a particular page. An example is when a page was already partially filled with code which was then executed and thus instrumented, after which new code is generated and placed on that page as well.
\end{itemize}

A better way is to make a copy of the accessed page, then mark it writable again and let the program resume its execution. This way, it can perform the changes it wanted to do itself. After a while, the instrumentor can compare the contents of the unprotected page and the the buffered copy to find the changes. So the question then becomes: when is this page checked for changes, how long will it be kept unprotected and how many pages will be kept unprotected at the same time. All parameters are important for performance, since keeping pages unprotected and checking them for changes requires both processing and memory resources. The when-factor is also important for correctness, as the modifications must be incorporated in the clone code before it is executed again.

On architectures with a weakly consistent memory model (such as the SPARC and PowerPC), the program must make its code changes permanent by using an instruction that synchronizes the instruction caches of all processors with the current memory contents. These instructions can be intercepted by the instrumentation engine and trigger a comparison of the current contents of a page with the previously buffered contents. On other architectures, heuristics have be used depending on the target application that one wants to instrument to get acceptable performance.

For example, when using the Sun JVM 1.4.0 running on a 80x86 machine under Linux, we compare the previously buffered contents of a page to the current contents whenever the thread that caused the protection fault does one of the following:

\begin{itemize}
\item{It performs a {\tt kill} system call.} This means the modifier thread is sending a signal to another thread, which may indicate that it has finished modifying the code and that it tells the other thread that it can continue.
\item{It executes a {\tt ret} or other instruction that requires a lookup to find the appropriate instrumentation code.} This is due to the fact that sometimes the modifying and executing threads synchronise using a spinlock. The assumption here is that before the modifying thread clears the spinlock, it will return from the modification routine, thus triggering a flush. Although this method is by no means a guarantee for correct behaviour in the general case, in our experience it always performs correctly in the context of instrumenting code generated by the Sun JVM 1.4.0.
\end{itemize}

The unprotected page is protected again when it has been checked {N} successive times without any changes having been made to it, or when another page has to be unprotected due to a protection fault. Note that this optimisation only really pays off in combination with only checking the page contents in the thread that caused the initial protection fault. The reason is that this ensures that the checking limit is not reached prematurely. Otherwise, the page is protected again too soon and a lot of extra page faults occur, nullifying any potential gains.

Finally, it is possible to vary the number of pages that are being kept unprotected at the same time. Possible strategies are keeping just one page unprotected for the whole program in order to minimize resources spent on buffering and comparing contents, keeping one page unprotected per thread, or keeping several pages unprotected per thread to reduce the amount of protection faults. Which strategy performs best depends on the cost of a page fault and the time necessary to do a page compare.

\section{Handling modifications}

Different code fragments in the clone are often interconnected by direct jumps. For example, when  -- while instrumenting -- we arrive at an instruction which was already instrumented before, we generate a direct jump to this previously instrumented version instead of instrumenting that code again. This not only improves efficiency, but it also makes the instrumentation of modified code much easier, since there is only one location in the clone we have to adapt in case of a code modification.

Because of these direct jump interconnections, merely generating an instrumented version of the modified code at a different location in the clone is not enough. Even if every lookup for the instrumented version of the code in that fragment returns one of the new addresses in the clone, the old code is still reachable via de direct jumps from other fragments. Removing the direct jumps and replacing them with lookups results in a severe slowdown.

Another solution would be keeping track of to which other fragments each fragment refers and adapting the direct jumps in case of changes. This requires a lot of bookkeeping however, and changing one fragment may result in a cascade effect, requiring a lot of additional changes elsewhere in the clone. For these reasons, we opted for the following three-part strategy.

The optimal way to handle the modifications, is to reinstrument the code in-place. This means that the previously instrumented version of the instructions in the clone are simply replaced by the new ones. This only works if the new code has the same length as (or is shorter than) the old code however, which is not always the case.

A second way to handle modifications can be applied when the instrumented version of the previous instruction at that location was larger than the size of an immediate jump. In this case, it is possible to overwrite the previous instrumented version with a jump to the new version. At the end of this new code, another jump can transfer control back to rest of the original instrumentation code.

Finally, if there is not enough room for an immediate jump, the last resort is filling the room originally occupied by the instrumented code with breakpoints. The instrumented version of the new code will simply be placed somewhere else in the code. Whenever the program then arrives at such a breakpoint, \diota{}'s exception handler is entered. This exception handler has access to the address where the breakpoint exception occurred, so it can use the translation table at the end of the block to look up the corresponding original program address. Next, it can lookup where the latest instrumented version of the code at that address is located and transfer control there.

\section{Experimental evaluation}

\begin{table*}[t]
\begin{center}
\begin{tabular}{|l||r|r|r|r|r|r|}
\hline Program & Normal & Instrumented & Slowdown & Relative \# of & Relative \# \\
name & execution (s) & execution (s) & ~ & protection faults & of lookups \\
\hline
FFT & 40.28 & 95.86 & 2.38 & 2305 & 409609 \\
\hline
MolDyn & 22.03 & 65.57 & 2.98 & 5105 & 423174 \\
\hline
SparseMatmult & 24.29 & 91.09 & 3.75 & 3751 & 874669 \\
\hline
HeapSort & 5.25 & 41.03 & 7.82 & 14779 & 1700553 \\
\hline
LUFact & 4.53 & 38.17 & 8.43 & 17402 & 1655753 \\
\hline
SearchBench & 23.92 & 429.10 & 17.94 & 8144 & 6337596 \\
\hline
Crypt & 8.91 & 175.15 & 19.66 & 12845 & 6696704 \\
\hline
RayTraceBench & 28.87 & 652.11 & 22.59 & 6611 & 8026878  \\
\hline
\end{tabular}
\end{center}
\caption{Test results for a number of sequential JavaGrande 2.0 benchmarks}
\label{tab:jvg}
\end{table*}

\subsection{General observations}

We evaluated the described techniques by implementing them in the \diota{} framework. The performance and correctness were verified using a number of tests from the JavaGrande~\cite{javagrande} benchmark, running under the Sun JVM 1.4.0 on a machine with two Intel Celeron processors clocked at 500MHz. The operating system was Redhat Linux 7.3 with version 2.4.19 of the Linux kernel.

Several practical implementation issues were encountered. The stock kernel that comes with Redhat Linux 7.3, which is based on version 2.4.9 of the Linux kernel, contains a number of flaws in the exception handling that cause it to lock up or reboot at random times when a lot of page protection exceptions occur. Another problem is that threads in general only have limited stack space and although \diota{} does not require very much, the exception frames together with \diota{}'s overhead were sometimes large enough to overflow the default stacks reserved by the instrumented programs. Therefore, at the start of the main program and at the start of every thread, we now instruct the kernel to execute signal handlers on an alternate stack.

\diota{}'s instrumentation engine is not re-entrant and as such is protected by locks. Since a thread can send a signal to another thread at any time, another problem we experienced was that sometimes a thread got a signal while it held the instrumentation lock. If the triggered signal handler was not yet instrumented at that point, \diota{} deadlocked when it tried to instrument this handler. Disabling all signals before acquiring a lock and re-enabling them afterwards solved this problem.

The problem with only parts of instructions being modified, which happens considerably more often than replacing whole instructions, was solved by adding a routine that finds the start of the instruction in which a certain address lies and starting the reinstrumentation from that address. Most modifications we observed were changes to the target addresses of direct calls.

The heuristics regarding only checking for changes in the thread that caused the initial unprotection of the page, reduced the slowdown caused by the instrumentation by 43\% relative to a strategy where pages are checked for changes every time a system call occurs and every time a lookup is performed, regardless of the involved threads. The limit on the number of checks done before a page is protected again (with {N} set between 3 and 5) provided an additional speed increase of 22\%.

A large part of the overhead stems from the fact that the ELF binary format (which is used by all modern Linux applications) permits code and data to be on the same page. The result is that once a code fragment on such a page has been instrumented, a lot of page faults occur and unnecessary comparisons have to be performed whenever such data is modified afterwards. A possible solution is not marking pages belonging to the ELF binary and the standard libraries loaded at startup time as read-only, but this could compromise the correctness of the described technique. However, it could be done if execution speed is of great concern and if one is certain that no such code will be modified during the execution.

\subsection{Test results}

Table \ref{tab:jvg} shows the measured timings when running a number of tests from sections 2 and 3 of the sequential part of the JavaGrande benchmark v2.0~\cite{javagrande}, all using the {S}ize{A} input set. The first column shows the name of the test program. The second and third columns show the used cpu time (as measured by the {\tt time} command line program, expressed in seconds) of an uninstrumented resp.\ instrumented execution, while the fourth column shows the resulting slowdown factor.

The fifth column contains the the amount of protection faults divided by the uninstrumented execution time, so it is an indication of the degree in which the program writes to pages that contain already executed code. The last column shows the number of lookups per second of uninstrumented execution time, where a lookup equals a trip to \diota{} via a trampoline to get the address of the instrumented code at the queried target address. The results have been sorted on the slowdown factor.

Regression analysis shows us that the overhead due to the lookups is nine times higher than that caused by the protection faults (and page compares, which are directly correlated with the number of protection faults, since every page is compared {N} times after is unprotected due to a fault). This means that the page protection technique has a quite low overhead and that most of the overhead can be attributed to the overhead of keeping the program under control.

The cause for the high cost of the lookups comes from the fact that the lookup table must be locked before it can be consulted, since it is shared among all threads. As mentioned before, we have to disable all signals before acquiring a lock since otherwise a deadlock might occur. Disabling and restoring signals is an extremely expensive operation under Linux, as both operations require a system call.

We have verified this by instrumenting a program that does not use use any signals nor self-modifying code using two different versions of \diota{}: one which does disable signals before acquiring a lock and one which does not. The first version is four times slower than the second one.

\section{Conclusions and future plans}

We have described a method which can be used to successfully instrument an important class of programs that use self-modifying code, specifically Java programs run in an environment that uses a {J}{i}{T}-compiler. The technique uses the hardware page protection mechanism present in the processor to detect modifications made to already instrumented code. Additionally, a number of optimisations have already been implemented to reduce the overhead, both by limiting the number of protection faults that occurs and the number of comparisons that must to be performed.

In the near future, a number of extra optimisations will be implemented, such as keeping more than one page unprotected at a time and the possibility to specify code regions that will not be modified, thus avoiding page protection faults caused by data and code being located on the same page. Additionally, we are also adapting the \diota{} framework in such a way that every thread gets its own clone and lookup table. This will greatly reduce the need for locking and disabling/restoring signals, which should also result in a significant speedup for the programs that perform a large number of lookups.

\newcommand{\etalchar}[1]{$^{#1}$}

\end{document}